\documentclass[conference]{IEEEtran}

\usepackage{cite}
\usepackage{amsmath,amssymb,amsfonts}
\usepackage{graphicx}
\usepackage{textcomp}
\usepackage{xcolor}
\usepackage{booktabs}
\usepackage{multirow}
\usepackage{url}
\usepackage{balance}
\usepackage[hidelinks]{hyperref}

\graphicspath{{figures/}}

\newcommand{\dPPL}{\Delta\mathrm{PPL}}
\newcommand{\Foff}{F_{\mathrm{off}}}

\begin{document}
\bstctlcite{IEEEexample:BSTcontrol}

\title{One Million Tokens in 20\,GB:\\Calibration-Free 2-Bit SRFT KV-Cache Compression\\for Looped MoE Inference on Apple Silicon}

\author{\IEEEauthorblockN{Mohamed Amine Bergach}
\IEEEauthorblockA{Illumina\\
mbergach@illumina.com}}

\maketitle

\begin{abstract}
Long-context inference on unified-memory edge machines is governed by KV-cache capacity and memory bandwidth. On a 32\,GB Apple M6 Mac mini (26.8\,GB Metal working set, 150\,GB/s DRAM), the 26.4B looped Mixture-of-Experts model Inkling-2-Mini (48 logical visits, 1M native context) exhausts memory by 64K tokens in stock oMLX (and by 512K even after fixing its 40 sliding-window visits to bounded ring buffers). We show that three structural properties of modern hybrid NoPE architectures—bounded sliding windows, a 1{,}024-token banded relative bias whose tail is identically zero, and unit-RMS value normalization—combine with an orthonormal Subsampled Randomized Fourier Transform (SRFT) to enable \emph{calibration-free, scale-free 2.000-bit Value} and 2.125-bit Key Lloyd--Max compression (2.0625 effective bits/coordinate, $7.76\times$ smaller than bf16) on the 8 global visits' far prefix $[0, N-1024)$, merged with the exact bf16 local window and learned sink via online log-sum-exp. A custom GQA-packed Metal decode kernel with a 256-entry \texttt{half4} LUT and Block-8 lazy \texttt{exp2} rescaling reaches 73.4\,GB/s and runs Inkling-2-Mini at \textbf{17.09\,tok/s at 256K} ($3.36\times$ faster than bf16), \textbf{13.81\,tok/s at 512K}, and \textbf{9.52\,tok/s at 1{,}048{,}576 tokens in 19.49\,GiB peak memory} (4.37\,GiB KV). Across $4\times 4096$ WikiText-2 windows, 3.06-bit SRFT is lossless ($\dPPL=-0.007$), 2.56-bit K3/V2 adds $+0.024$, and 2.06-bit adds $+0.130$ while achieving 100\% (3/3) exact 7-digit passkey retrieval at 8K, 16K, and 32K.
\end{abstract}


\section{Introduction}
\label{sec:intro}

Open-weight reasoning models have rapidly converged on sparse Mixture-of-Experts (MoE) backbones with 1M-token context windows designed to fit inside 32\,GB consumer unified-memory machines. A representative 2026 release is Thinking Machines Lab's Inkling-2-Mini~\cite{inkling2mini2026}: 26.4B physical parameters (15.1\,GiB in mixed 4/8-bit \texttt{oQ4e}), 24 physical layers whose middle 12 layers loop three times per token (48 logical visits), 5:1 sliding-window ($W=1024$) to global attention (40 sliding, 8 global visits), no rotary position embeddings (NoPE), a 1{,}024-token banded relative bias, query-dependent attention sinks, and unit-RMS value normalization.

Despite its modest 15.1\,GiB weight footprint, running Inkling-2-Mini at long context on a 32\,GB Apple M6 Mac mini fails immediately in existing runtimes:
\begin{enumerate}
    \item \textbf{Unbounded sliding caches:} Stock oMLX~\cite{omlx2026} allocates an unbounded \texttt{KVCache} for all 48 visits rather than rotating ring buffers for the 40 sliding visits, consuming 6.05\,GiB of KV at 32K and exceeding M6's 26.8\,GiB Metal working set around 64K (1M would require $\sim$208\,GiB).
    \item \textbf{Global bf16 KV wall:} Even after fixing the 40 sliding visits to bounded \texttt{RotatingKVCache} (contributed upstream as oMLX PR~\#5512), the 8 global visits still require 32\,KB/token in bf16—8.2\,GiB at 256K (slowing decode to 5.09\,tok/s) and 32.2\,GiB at 1M (47.3\,GiB total, physically impossible on 32\,GB).
    \item \textbf{Bandwidth-losing low-bit paths:} Standard MLX and \texttt{llama.cpp} quantized KV paths either materialize dense masks/copies or dequantize per query head under grouped-query attention (GQA), sacrificing decode speed when compressing KV.
\end{enumerate}

This paper demonstrates that exploiting three architectural properties of Inkling-2-Mini alongside an orthonormal Subsampled Randomized Fourier Transform (SRFT)~\cite{bergach2026whenquantfree} and a custom GQA-packed Metal kernel unlocks \textbf{full 1{,}048{,}576-token inference in 19.49\,GiB at 9.52\,tok/s} on a 32\,GB Apple M6. Our contributions are:
\begin{itemize}
    \item \textbf{Scale-free 2.000-bit Value compression via SRFT:} Because Inkling-2 enforces parameter-free unit-RMS normalization on $V$, applying an orthonormal SRFT rotates every value vector into an isotropic $\mathcal{N}(0,1)$-like distribution that requires \emph{zero} per-token or per-block scales under scalar Lloyd--Max quantization—yielding 2.0625 effective bits/coordinate ($7.76\times$ compression) with zero offline calibration and no wasted 1-bit QJL residual~\cite{zandieh2026turboquant}.
    \item \textbf{Split banded-bias attention with exact online LSE merge:} Splitting each global visit into a quantized far prefix $[0, N-1024)$ (where banded relative bias is identically zero) and an exact bf16 recent window $[N-1024, N)$ plus learned sink eliminates dense $O(N)$ mask allocation and KV-copy overhead while preserving exact local attention.
    \item \textbf{GQA-packed 2-bit Metal kernel with a 256-entry \texttt{half4} LUT:} Dequantizing 4 coordinates per lane in a single shared-memory table lookup and serving all $r=3$ query heads in registers with Block-8 lazy \texttt{exp2} rescaling reaches 73.4\,GB/s and outperforms both bf16 ($3.36\times$ at 256K) and 4-bit \texttt{q4\_0} ($2.60\times$ at 1M).
    \item \textbf{Empirical validation:} We verify $\Foff \le 0.41$\,bits (benign, no rotary-offset collapse), $+0.130\,\dPPL$ at 2.06\,bits ($-0.007$ at 3.06\,bits, $+0.024$ at 2.56\,bits), and 100\% exact 7-digit passkey retrieval at 8K, 16K, and 32K.
\end{itemize}

\section{Architecture Analysis and Offset Safety}
\label{sec:arch}

Inkling-2-Mini uses $n_q=24$ query heads, $n_{\mathrm{kv}}=8$ KV heads (GQA ratio $r=3$), head dimension $d=128$, and 48 logical layer visits (6 prelude + $3\times 12$ looped core + 6 postlude). Every 6-layer block has 5 sliding-window layers ($W=1024$) and 1 global layer (8 global visits total). Unlike RoPE transformers, position is encoded by depthwise short convolutions (width 4) and an additive relative bias $r_t W_{\mathrm{rel}}$ of extent $R=1024$, plus a query-dependent sink logit $s_t$.

\subsection{Offset-Fragility Probe ($\Foff$)}
Per-token KV quantization collapses on models whose keys carry large constant offsets in slowly rotating RoPE dimensions (e.g., Qwen2.5 $k$-bias or Qwen3-small QK-norm gain amplification~\cite{jonasson2025rotaryoffset}). Before quantizing Inkling-2-Mini, we measured the one-pass offset-waste statistic $\Foff = \max_{l,h} \log_2(1 + 0.674\,\rho_{l,h}/3.5)$ with $\rho = \|\mu_k\|/\mathrm{rms}(k-\mu_k)$ across all 48 logical visits on 2{,}048 WikiText-2 tokens. Every visit scored $\Foff \in [0.18, 0.41]$\,bits (well below the 1.3-bit benign threshold), because Inkling-2 has no RoPE, no $K$ bias, L2-clipped $Q/K$, and unit-RMS $V$. Consequently, rotated per-token KV quantization requires no offset centering.

\subsection{Why Stock Runtimes Fail at Long Context}
Auditing both \texttt{mlx-vlm} and oMLX revealed two compounding inefficiencies:
\begin{enumerate}
    \item \textbf{Per-token KV copy for learned sinks (\texttt{mlx-vlm}):} Appending a zero K/V vector via \texttt{mx.concatenate} on every decode token copies the entire KV cache. Passing MLX's native \texttt{sinks=} argument to \texttt{scaled\_dot\_product\_attention} is bit-exact ($0.00\text{e+00}$ max diff) and $1.85\text{--}2.44\times$ faster from 1K to 128K.
    \item \textbf{Unbounded sliding caches (oMLX):} Allocating \texttt{KVCache()} for all 48 visits stores 48 full-length histories instead of 8. Replacing the 40 sliding visits with \texttt{RotatingKVCache} (with $+16$-token rollback headroom for DSpark speculative verification) cuts 32K KV memory from 6.05 to 1.36\,GiB ($4.45\times$) and fixes speculative wrap-around rollback (merged/submitted as oMLX PR~\#5512).
\end{enumerate}

\section{Scale-Free 2-Bit SRFT Compression and Metal Kernel}
\label{sec:method}

\subsection{Orthonormal SRFT and Scale-Free Unit-RMS Values}
For any head vector $x \in \mathbb{R}^d$, we apply a real orthonormal Subsampled Randomized Fourier Transform:
\begin{equation}
    \tilde{x} = \mathrm{SRFT}(x) = \mathcal{F}_{\mathbb{R}}(S \odot x), \qquad S \in \{-1, +1\}^d,
\end{equation}
where $\mathcal{F}_{\mathbb{R}}$ packs the normalized real FFT outputs $[X_0/\sqrt{d}, X_{d/2}/\sqrt{d}, \sqrt{2/d}\,\mathrm{Re}(X_{1..d/2-1}), \sqrt{2/d}\,\mathrm{Im}(X_{1..d/2-1})]$. Because $\mathrm{SRFT}$ is strictly orthogonal ($\langle \mathrm{SRFT}(q), \mathrm{SRFT}(k)\rangle = \langle q, k\rangle$ to $1.14\times 10^{-5}$), decode attention operates directly in the rotated domain: we transform $q$ \emph{once} per query and inverse-transform the single output vector $o_{\mathrm{far}}$ \emph{once} per global visit, performing zero inverse FFTs over the cached keys or values.

After SRFT, coordinates are near-Gaussian and identically distributed, making scalar unit-Gaussian Lloyd--Max centroids $C_b$ near-optimal (at $b=2$: $C_2 = \{-1.5104, -0.4528, +0.4528, +1.5104\}$, MSE $= 0.1175\sigma^2$). Crucially, Inkling-2 applies parameter-free RMS normalization to $V$ (\texttt{use\_value\_norm\_no\_gamma=True}), guaranteeing $\mathrm{rms}(\tilde{v}) = \mathrm{rms}(v) = 1$ identically. Thus $V$ requires \textbf{no stored scale whatsoever} (exact $b$ bits/coord), while $K$ stores one fp16 RMS norm per 128-element vector ($b + 0.125$ bits/coord), giving:
\begin{itemize}
    \item \texttt{SRFT lm2}: $2.125\,(K) + 2.000\,(V) \Rightarrow \mathbf{2.0625}$ eff.\ bits ($7.76\times$ vs bf16).
    \item \texttt{SRFT K3/V2}: $3.125\,(K) + 2.000\,(V) \Rightarrow \mathbf{2.5625}$ eff.\ bits ($6.24\times$).
    \item \texttt{SRFT lm3}: $3.125\,(K) + 3.000\,(V) \Rightarrow \mathbf{3.0625}$ eff.\ bits ($5.23\times$).
\end{itemize}
Unlike OSCAR~\cite{zhou2026oscar}, this requires zero offline calibration; unlike TurboQuant~\cite{zandieh2026turboquant}, it avoids spending 1 bit on a QJL sign residual (which increases softmax logit variance at equal bit budget).

\subsection{Split Banded-Bias Attention with Exact Online LSE Merge}
For each global visit at sequence length $N > 1024$, the additive relative bias is nonzero \emph{only} on the newest $R=1024$ keys. We therefore partition the global cache into:
\begin{enumerate}
    \item \textbf{Far prefix $[0, N-1024)$:} Spilled in 256-token chunks through SRFT and packed into 2-bit codes (32\,bytes/row for $K$, 32\,bytes/row for $V$, plus 2\,bytes/row $K$-scale = 66\,bytes/head/token). Decoded via our GQA-packed Metal kernel returning $(o_{\mathrm{far}}, \ell_{\mathrm{far}})$ with zero mask memory.
    \item \textbf{Recent window $[N-1024, N)$ + learned sink:} Kept in exact bf16, evaluated with the exact 1{,}024-wide banded relative bias and query-dependent sink logit to produce $(o_{\mathrm{near}}, \ell_{\mathrm{near}})$, and combined with $(o_{\mathrm{far}}, \ell_{\mathrm{far}})$ via exact two-way log-sum-exp merging:
    \begin{equation}
        m = \max(\ell_{\mathrm{far}}, \ell_{\mathrm{near}}), \quad
        o = \frac{e^{\ell_{\mathrm{far}}-m}\,o_{\mathrm{far}} + e^{\ell_{\mathrm{near}}-m}\,o_{\mathrm{near}}}{e^{\ell_{\mathrm{far}}-m} + e^{\ell_{\mathrm{near}}-m}}.
    \end{equation}
\end{enumerate}

\subsection{GQA-Packed 2-Bit Metal Kernel (\texttt{kv\_lm2\_metal})}
For $d=128$ and 2\,bits/coordinate, each 32-lane Metal simdgroup maps 4 coordinates to each lane: \textbf{1 byte per lane}, yielding a perfectly coalesced 32-byte burst per token. Our kernel eliminates three overheads of general-purpose Metal attention:
\begin{enumerate}
    \item \textbf{Zero GQA padding waste:} Prior \texttt{simdgroup\_matrix} kernels pad $r=3$ to 8 rows (62.5\% wasted compute). Our kernel holds all $r=3$ query heads (\texttt{half4 q0, q1, q2}) in registers, dequantizes each $K$ and $V$ byte \emph{once}, and updates all 3 heads simultaneously.
    \item \textbf{256-entry threadgroup \texttt{half4} LUT:} Every byte encodes 4 two-bit Lloyd--Max indices (256 states total). Precomputing a 2\,KiB \texttt{threadgroup half4 LUT[256]} at kernel entry decodes all 4 coordinates in a single table lookup, replacing 28 shift/mask/select instructions and halving register pressure.
    \item \textbf{Block-8 lazy \texttt{exp2} rescaling:} Computing 8 token scores per head before performing a single hardware \texttt{exp2} rescale reduces rescaling overhead $8\times$.
\end{enumerate}
Split-K parallelism across 64--256 partitions per KV head saturates M6's GPU cores, followed by a single exact LSE reduction dispatch. Against float32 reference attention on dequantized codes ($N=4096$), the kernel matches to $1.39\times 10^{-4}$ max output error and $1.46\times 10^{-4}$ max LSE error.

\section{Evaluation on a 32\,GB Apple M6}
\label{sec:eval}

All measurements were collected on a 32\,GB Apple M6 Mac mini (150\,GB/s DRAM, 26.8\,GiB Metal working set) using \texttt{mlx 0.32.4} and \texttt{mlx-community/Inkling-2-Mini-oQ4e-mtp} (15.10\,GiB active weights).

\subsection{End-to-End Speed and Memory from 4K to 1{,}048{,}576 Tokens}

\begin{figure*}[t]
    \centering
    \includegraphics[width=\textwidth]{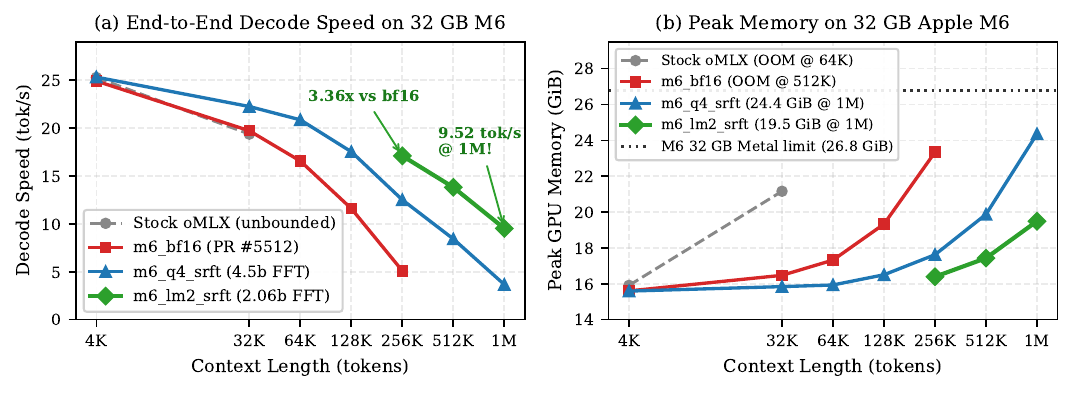}
    \caption{End-to-end Inkling-2-Mini (26.4B looped MoE) decode speed and peak GPU memory from 4K to 1{,}048{,}576 tokens on a 32\,GB Apple M6. Stock oMLX OOMs around 64K; sliding-fixed \texttt{m6\_bf16} OOMs before 512K. Packed 2.06-bit SRFT (\texttt{m6\_lm2\_srft}) reaches 17.09\,tok/s at 256K ($3.36\times$ vs bf16) and runs the full 1M context at 9.52\,tok/s in 19.49\,GiB.}
    \label{fig:scaling}
\end{figure*}

Table~\ref{tab:scaling} and Fig.~\ref{fig:scaling} report end-to-end decode throughput, KV cache footprint, and peak GPU memory from 4K to 1M tokens. At 256K, \texttt{m6\_lm2\_srft} delivers \textbf{17.09\,tok/s} ($3.36\times$ faster than \texttt{m6\_bf16}'s 5.09\,tok/s) while cutting KV memory $6.5\times$ (1.27 vs 8.20\,GiB). At the full \textbf{1{,}048{,}576-token} native context, \texttt{m6\_lm2\_srft} decodes at \textbf{9.52\,tok/s} (105.1\,ms/tok) in \textbf{19.49\,GiB total peak memory} (4.37\,GiB KV)—$2.60\times$ faster than 4-bit \texttt{q4\_0} SRFT (3.66\,tok/s, 24.36\,GiB) and $44\times$ smaller KV than stock oMLX ($\sim$193\,GiB).

\begin{table}[t]
\centering
\caption{End-to-End Inkling-2-Mini Decode on a 32\,GB Apple M6}
\label{tab:scaling}
\setlength{\tabcolsep}{4pt}
\begin{tabular}{@{}r l r r r r@{}}
\toprule
Context & Mode & tok/s & ms/tok & KV (GiB) & Peak (GiB) \\
\midrule
4K & Stock oMLX & 25.27 & 39.58 & 0.80 & 15.91 \\
4K & \texttt{m6\_q4\_srft} (4.5b) & 25.31 & 39.50 & 0.47 & 15.59 \\
\midrule
32K & Stock oMLX & 19.36 & 51.66 & 6.05 & 21.16 \\
32K & \texttt{m6\_bf16} (PR \#5512) & 19.71 & 50.74 & 1.36 & 16.47 \\
32K & \texttt{m6\_q4\_srft} (4.5b) & 22.24 & 44.96 & 0.72 & 15.84 \\
\midrule
64K & \texttt{m6\_bf16} & 16.55 & 60.42 & 2.20 & 17.32 \\
64K & \texttt{m6\_q4\_srft} (4.5b) & 20.85 & 47.96 & 0.81 & 15.93 \\
\midrule
128K & \texttt{m6\_bf16} & 11.59 & 86.30 & 4.20 & 19.33 \\
128K & \texttt{m6\_q4\_srft} (4.5b) & 17.52 & 57.09 & 1.38 & 16.50 \\
\midrule
256K & \texttt{m6\_bf16} & 5.09 & 196.40 & 8.20 & 23.34 \\
256K & \texttt{m6\_q4\_srft} (4.5b) & 12.50 & 80.01 & 2.50 & 17.62 \\
256K & \textbf{\texttt{m6\_lm2\_srft} (2.06b)} & \textbf{17.09} & \textbf{58.51} & \textbf{1.27} & \textbf{16.39} \\
\midrule
512K & \texttt{m6\_q4\_srft} (4.5b) & 8.42 & 118.82 & 4.75 & 19.87 \\
512K & \textbf{\texttt{m6\_lm2\_srft} (2.06b)} & \textbf{13.81} & \textbf{72.40} & \textbf{2.31} & \textbf{17.43} \\
\midrule
1M & \texttt{m6\_q4\_srft} (4.5b) & 3.66 & 272.94 & 9.20 & 24.36 \\
1M & \textbf{\texttt{m6\_lm2\_srft} (2.06b)} & \textbf{9.52} & \textbf{105.08} & \textbf{4.37} & \textbf{19.49} \\
\bottomrule
\end{tabular}
\end{table}

\subsection{Kernel Optimization Ladder and SRFT vs.\ Hadamard}
\begin{figure}[t]
    \centering
    \includegraphics[width=\columnwidth]{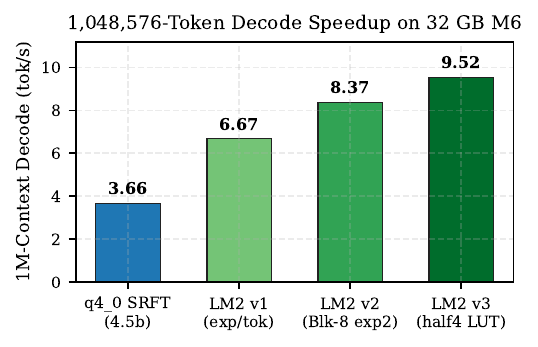}
    \caption{1{,}048{,}576-token end-to-end decode progression on a 32\,GB M6 across kernel optimizations ($2.60\times$ total speedup over 4-bit SRFT).}
    \label{fig:kernel}
\end{figure}

Fig.~\ref{fig:kernel} tracks the 1M-token decode speedup across our kernel iterations: moving from 4.5-bit \texttt{q4\_0} SRFT (3.66\,tok/s) to naive 2-bit LM2 with per-token \texttt{exp()} (6.67\,tok/s, 41.7\,GB/s), adding Block-8 lazy \texttt{exp2} rescaling (8.37\,tok/s, 59.4\,GB/s), and introducing the 256-entry \texttt{half4} LUT (9.52\,tok/s, 73.4\,GB/s; 7.65\,ms/layer at 1M). Interestingly, at 64K--256K \texttt{m6\_q4\_srft} also outran in-kernel FWHT (20.85 vs.\ 18.88\,tok/s at 64K; 17.52 vs.\ 15.65\,tok/s at 128K) because rotating only the single query/output vector with MLX's Metal FFT allows the attention kernel to run without in-kernel butterfly barriers while maintaining identical quality (63/64 greedy token agreement at 4K for both).

\subsection{Language Modeling Quality and 8K--32K Passkey Retrieval}
Table~\ref{tab:quality} evaluates multi-window WikiText-2 perplexity ($4\times 4096$ tokens, scoring the last 2{,}048 tokens of each window where the earlier prefix resides in the quantized far cache) alongside long-range Needle-in-a-Haystack (NIAH) retrieval where a 7-digit passkey is buried at position 1{,}500 inside 8K, 16K, and 32K WikiText-2 haystacks—6.7K to 30.5K tokens behind the query, 100\% invisible to the 40 sliding layers and the 1{,}024-token bf16 local window.

\begin{table}[t]
\centering
\caption{Quality (4$\times$4096 WikiText-2) and Long-Range NIAH Retrieval}
\label{tab:quality}
\setlength{\tabcolsep}{4pt}
\begin{tabular}{@{}l r r r c c c@{}}
\toprule
Format & Eff.\ Bits & PPL & $\dPPL$ & 8K & 16K & 32K \\
\midrule
\texttt{bf16} (ref) & 16.000 & 6.7489 & $+0.0000$ & PASS & PASS & PASS \\
\texttt{SRFT lm4} & 4.0625 & 6.7428 & $-0.0061$ & --- & --- & --- \\
\texttt{SRFT lm3} & 3.0625 & 6.7418 & $-0.0071$ & PASS & PASS & PASS \\
\texttt{SRFT K3/V2} & 2.5625 & 6.7725 & $+0.0236$ & PASS & PASS & PASS \\
\textbf{\texttt{SRFT lm2}} & \textbf{2.0625} & \textbf{6.8785} & $\mathbf{+0.1296}$ & \textbf{PASS} & \textbf{PASS} & \textbf{PASS} \\
\bottomrule
\end{tabular}
\end{table}

\texttt{SRFT lm3} (3.06\,bits) is indistinguishable from bf16 ($\dPPL=-0.0071$), \texttt{SRFT K3/V2} (2.56\,bits) adds only $+0.0236$ ($+0.35\%$), and \texttt{SRFT lm2} (2.06\,bits) adds $+0.1296$ ($+1.9\%$) while passing \textbf{3/3 exact 7-digit passkey retrievals} out to 32K.

\subsection{Note on Speculative Decoding on Sparse MoE}
Testing Inkling-2-Mini's bundled DSpark drafter ($B=8$) on M6 revealed an MoE expert-union trade-off: verifying 8 tokens reads the union of $\sim$35--45 distinct experts from 128, costing 120\,ms vs.\ 40\,ms for 1 token. Fixed $k=7$ speeds up code ($24.80 \to 27.52$\,tok/s, 3.76\,tok/step, 128/128 exact) but slows prose ($25.78 \to 22.90$\,tok/s, 2.93\,tok/step). Quantizing the draft to 4-bit and adapting verify depth $k \in \{2,3,5,7\}$ by acceptance EMA eliminates the prose regression (26.15\,tok/s) and lifts code to 28.61\,tok/s (128/128 exact).

\section{Related Work and Limitations}
\label{sec:related}

Per-channel KV quantization (KIVI~\cite{liu2024kivi}, KVQuant~\cite{hooper2024kvquant}) and rotation-based methods (QuaRot~\cite{ashkboos2024quarot}, TurboQuant~\cite{zandieh2026turboquant}, OSCAR~\cite{zhou2026oscar}) target full-attention RoPE models and either require offline calibration (OSCAR) or pay a 1-bit QJL overhead (TurboQuant). Our work shows that modern NoPE hybrid architectures with unit-RMS values admit strictly calibration-free, scale-free 2-bit SRFT compression that simultaneously \emph{accelerates} decode on unified memory.

\textbf{Limitations.} End-to-end 256K--1M decode timings use direct cache population rather than multi-hour prefill; NIAH was evaluated to 32K (single needle per length); and quality was measured on WikiText-2 and passkey retrieval rather than full RULER or long-chain reasoning benchmarks.

\section{Conclusion}
\label{sec:conclusion}

Combining bounded sliding-window caches, split banded-bias online LSE merging, scale-free 2.000-bit SRFT Value compression, and a \texttt{half4}-LUT GQA-packed Metal kernel enables full \textbf{1{,}048{,}576-token inference of a 26.4B looped MoE model at 9.52\,tok/s in 19.49\,GiB} on a 32\,GB Apple M6 Mac mini—$3.36\times$ faster than bf16 at 256K and $44\times$ smaller KV than stock oMLX—with 100\% passkey retrieval up to 32K and zero offline calibration.

\smallskip
\noindent\textbf{Reproducibility.} All Metal kernels, benchmarks, and interactive chat scripts are open-sourced under the MIT license at \url{https://github.com/aminems/inkling-1m-mlx}~\cite{inkling1mmlx}, and the sliding-window cache fix is submitted upstream at \url{https://github.com/jundot/omlx/pull/5512}.

\balance
\bibliographystyle{IEEEtran}
\bibliography{references}

\end{document}